\documentclass[usenatbib]{mn2e}
\usepackage{times,url}
\usepackage{graphicx,amsmath,amssymb}

\pubyear{2009}
\author[Bagla, Kulkarni \& Padmanabhan]{J.~S.~Bagla$^1$, Girish Kulkarni$^1$, 
  T.~Padmanabhan$^2$\\$^1$Harish-Chandra Research Institute, Chhatnag
  Road, Jhunsi, Allahabad 211019 India.\\
  $^2$Inter University Centre for Astronomy and
  Astrophysics, Post Bag 4, Ganeshkhind, Pune 411007 India.\\
  E-mail: girish@hri.res.in, jasjeet@hri.res.in, nabhan@iucaa.ernet.in} 
\title[Constraints on Early Star Formation]{Metal Enrichment and
  Reionization Constraints on Early Star Formation}
\label{firstpage}
\pagerange{\pageref{firstpage}--\pageref{lastpage}}
\begin{document}

\maketitle

\begin{abstract}
  The epoch of reionization and formation of first stars are
  inter-linked topics that are of considerable interest.  
  We use a simplified approach for studying formation of stars in
  collapsed haloes, and the resulting ionization of the inter-galactic
  medium (IGM). 
  We consider a set of $\Lambda$CDM models allowed by observations of
  CMB temperature and polarization anisotropies for this study.  
  We constrain parameters related to star formation with the help of
  observations.
  We constrain subsets of these parameters independently by using the
  observed metallicity of the inter-galactic medium at $z \sim 5$ and
  the requirement that the Thomson scattering optical depth due to an
  ionized IGM as determined for the model from CMB observations be
  reproduced.
  We consider a range of initial metalicities for star forming gas,
  and some variations of the initial mass function of stars.  
  We find that a ``normal'' initial mass function (IMF) may satisfy
  these two constraints with a raised efficiency of star formation as compared
  to that seen in the local universe.
  Observations require a significant fraction of metals to
  escape from haloes to the IGM.
  We can also place constraints on the ratio of escape fraction for
  metals and ionizing photons, we find that this ratio is of order unity for
  most models.
  This highlights the importance of using the constraints arising from
  enrichment of the inter-galactic medium.
  Ultra-high mass stars or AGNs may not simplify models of
  reionization in that these may produce more ionizing photons but
  these do not contribute to production of metals and hence these help
  in reducing only the escape fraction for ionizing photons.
  However, suppression of very low mass stars is helpful in that it
  increases the production of metals as well as ionizing photons and
  hence leads to a reduction in both escape fractions.
  Such a change is also warranted by observations of metal poor halo stars in
  the Galaxy.
  We also discuss correlations in parameters like the efficiency
  of star formation and the two escape fractions with cosmological
  parameters. 
\end{abstract}


\section{Introduction}

Large scale structures in the universe like galaxies and clusters of
galaxies are believed to have formed by gravitational amplification of
small perturbations \citep{1980lssu.book.....P, 1989RvMP...61..185S,
1999coph.book.....P, 2002tagc.book.....P, 2002PhR...367....1B}.  
Much of the matter in galaxies and clusters of galaxies is the so
called dark matter that is believed to be weakly interacting and
non-relativistic \citep{1987ARA&A..25..425T, 2008arXiv0803.0547K}.
Dark matter responds mainly to gravitational forces, and by virtue of
larger density than baryonic matter, assembly of matter into haloes
and large scale structure is driven by gravitational instability of
initial perturbations.  
Galaxies are believed to form when gas in highly over-dense haloes
cools and collapses to form stars in significant numbers
\citep{1953ApJ...118..513H, 1977MNRAS.179..541R, 1977ApJ...211..638S,
1977ApJ...215..483B}.  
The formation of first stars \citep{2007ARA&A..45..565M,
2007ARA&A..45..481Z, 2004ARA&A..42...79B} in turn leads to emission of
UV radiation that starts to ionize the inter-galactic medium (IGM).
The period of transition of the IGM from a completely neutral to a
completely ionized state is known as the epoch of reionization (EoR),
e.g., see \citet{2001ARA&A..39...19L}.

The study of EoR has been an active area of research in recent years.
Theoretical ideas about the reionization history have been constrained
by a variety of observations \citep{2006ARA&A..44..415F}.  
For example, observations of Gunn-Peterson troughs in AGN spectra at
$z \sim 6$ \citep{2001AJ....122.2850B, 2006AJ....132..117F} indicate
that the process of reionization was nearly complete by that redshift.
Bounds on luminosity function of Ly$\alpha$ galaxies at high redshifts
\citep{2004ApJ...617L...5M, 2005ApJ...619...12S,2008ApJ...686..230B} also
constrain the EoR.  
These bounds are consistent with the conclusion that the IGM was
completely ionized by $z \sim 6$. 
Furthermore, Thomson scattering by free electrons in the IGM of the
anisotropic photon distribution that constitutes the CMB leaves a
signature in its temperature and polarization anisotropy. 
The optical depth due to this scattering \citep{2003ApJS..148..175S,
2008arXiv0803.0586D} is another probe of EoR. 
WMAP five-year data indicate a value of $\tau\sim 0.084\pm 0.016$ that
corresponds to $z\sim 10$ as the redshift for instantaneous reionization.
Realistic scenarios however predict a protracted EoR where the process of
ionization starts around $z \sim 15$ and ends by $z \sim 6$.

Many possible sources of ionizing radiation have been considered,
although stellar sources are believed to be the most plausible
candidates \citep{2004ApJ...600L...1Y,2006MNRAS.369..825S,2007MNRAS.380L...6C}.
The AGN density goes down more
rapidly for $z>3$ than the density of star-forming galaxies
\citep{1990ApJ...350....1M, 1999ApJ...514..648M, 2001ApJ...549L.151H} and
therefore AGNs are not expected to contribute significantly to the ionizing
radiation. 
X-rays from low mass quasars, X-ray binaries and supernova remnants
are constrained by the soft x-ray background observed today
\citep{2004ApJ...613..646D}.
Particle decays can also play only a minor role in reionization
\citep{1998ApJS..114...37B, 2005MNRAS.364....2M}.
In the present work, we assume that it was radiation from early stars
that ionized the IGM and ignore other possibilities.

The total photon emissivity of early stars is poorly known.
Studies of reionization typically use observations with semi-analytic
models of heating and ionization of the IGM where efficiency of star
formation, evolution of star formation rates, the number of ionizing
photons emitted per baryon in stars, etc.\ are parameterized in some
manner \citep{2000ApJ...534..507C}.
Given the complexity of most of these approaches, and the number of
parameters, it is often impractical to scan the parameter space. 
The main lesson we learn from these studies is that we may not require
extraordinary physical processes in order to satisfy available observational 
constraints of reionization. 
We make an attempt to simplify modeling of star formation and other
astrophysical aspects of the problem.  
This allows us to reduce the number of free parameters in this sector
while retaining many significant astrophysical relationships.
Statements can then be made about quantities like the escape
fraction of UV photons, $f_{\mathrm{esc}, \gamma}$ and their correlations with
the cosmological parameters. 

Large uncertainties exist in parameters related to early star
formation, e.g.\ the escape fraction of UV photons, $f_{\mathrm{esc},
\gamma}$, the initial mass function (IMF) and the efficiency of star
formation, $f_\ast$ \citep{2004MNRAS.355..374B}.
It has been suggested in the literature that a top-heavy IMF with very
massive stars is not necessarily favored to satisfy the reionization
and metal enrichment constraints \citep{2006ApJ...647..773D,
2003ApJ...594L...1V}.

In this work we assume that early star formation happens predominantly
during formation and major mergers of haloes, and occurs as a short lived
burst.   
Photon emissivities and metal yields can then be calculated using
population synthesis models for different IMFs
\citep{1999ApJS..123....3L, 2003MNRAS.344.1000B}.
We then test if these scenarios generate enough photons to ionize the
universe by $z\sim 6$ by comparing with the observed Thomson
scattering optical depth.
We also require the models to satisfy constraints arising from the
observations of the metal content of the IGM. 
This constrains the amount of processed elements that escape from
the ISM to the IGM. 
We can use models for outflows as a guide and put constraints on the
efficiency of star formation, or use ``reasonable'' values of the
efficiency of star formation to constrain the metal escape fraction. 
We can also combine the two constraints to scale out efficiency of
star formation.
The questions we address in this work are:
\begin{itemize}
\item
To combine constraints of enrichment of the inter-galactic medium with
observations of reionization, and check whether the extra information can
provide constraints on the initial mass function during early star formation. 
\item
Can the combined constraints be used to make useful statements with regard to
other potential sources of ionization?
\item
Is there any correlation between the parameters that describe star formation
and cosmological parameters?
\end{itemize}

Relevant observations are described in
\S\ref{observation}.
The reionization model that we use is discussed in \S\ref{reionization}. 
Results and discussion appear in \S\ref{results} and \S{5}. 

\begin{table*}
\begin{center}
\begin{tabular}{||c|l|c|c|c|c|c|c|}
\hline
\hline
 & IMF & $M_\mathrm{low}/M_\odot$ & $M_\mathrm{high}/M_\odot$ &
$Z_\mathrm{input}$ & 
$N_\gamma$ & $p$ & $N_\gamma f_{\mathrm{esc}, \gamma} f_\ast$\\
\hline
1. & Kroupa & $0.1$ & $100$ & $0.0004$ & $6804$ & $0.0123$ & 50.0\\
2. & Kroupa & $0.5$ & $100$ & $0.0004$ & $9280$ & $0.0167$ & 50.0\\
3. & Kroupa & $0.1$ & $100$ & $0.001$ & $6297$ & $0.0159$ & 50.0\\
4. & Salpeter & $1$ & $100$ & $0.001$ & $11237$ & $0.0283$ & 50.0\\
5. & Kroupa & $0.1$ & $100$ & $0.02$ & $3996$ & $0.0261$ & 50.2\\
\hline
\end{tabular}
\end{center}
\caption{Various IMFs used in this study are summarized here.
  $M_\mathrm{low}$ and 
  $M_\mathrm{high}$ are the lower and upper mass cut-offs for the IMFs
  respectively.  $Z_\mathrm{input}$ denotes the metallicity of the gas
  from which stars are formed. $N_\gamma$ is the number of ionizing
  photons produced per baryon in stars and $p$ is the metal yield per
  baryon in them.  $N_\gamma$ and $p$ are obtained using population
  synthesis models.  The last column lists the value of $N_\gamma
  f_{\mathrm{esc},\gamma}f_\ast$ for the WMAP5 best-fit model with the
  corresponding IMF. This quantity is proportional to the number of
  photons escaping into the IGM for every baryon inside a collapse
  halo.}
\label{sfmodels}
\end{table*}

\section{Observations and Models}
\label{observation}

The IGM occupies most of the space and a substantial amount of matter in the
Universe.   
Indeed, it is believed that at least half of the baryons in the
universe are in the IGM. 
Thus it is not surprising that the observations of IGM dominate when
we discuss constraints on models of reionization. 
We introduce observations that are used for constraining models in
this paper.

Observational constraints that we use are essentially two: metallicity
of the IGM as seen in quasar absorption systems, and the Thomson
scattering optical depth from the cosmic microwave background. 
A third observation that can potentially be used is that of the
cosmic stellar matter density, which we model in a manner explained
below. 
We do not discuss constraints arising from this type of observations as the
observational bounds on models are not very strong at present.
It is not possible to use other observational constraints
\citep{2007MNRAS.380L...6C} in the global averaged model discussed here.

Observations of galaxies at high redshifts can be used to infer the
density of stellar matter, $\rho_*$, at those redshifts. 
One way of estimating this quantity is to use the luminosity
function of galaxies in various wavelength passbands and combine these
with population synthesis models and an assumed IMF
\citep{1996MNRAS.283.1388M, 1996ApJ...460L...1L, 2007ApJ...670..928B}.  
We do not use these observations here as current observations do not provide
sufficiently strong constraints. 
This is expected to change in coming years with better observations.

Instead in our work we assume that stars form only in virialised
haloes.  We take the minimum mass of star-forming haloes to be
$10^8$~M$_\odot$, this is a proxy for haloes with a virial temperature of
$10^4$~K. 
We do not take into account star formation in haloes of around
$10^6$~M$_\odot$ that is aided by molecular cooling
\citep{1997ApJ...474....1T}. 
We also ignore the effects of feedback that raise the Jeans mass to
around $10^9$~M$_\odot$\footnote{We have checked that including the effects of
radiative feedback increases the required star formation efficiency by around
$20\%$.}. 
We can then obtain metal and photon yields by using population
synthesis models: we use the {\scshape starburst99} code
\citep{1999ApJS..123....3L,2005ApJ...621..695V}. 
This assumption allows us to connect the rate of change of stellar
mass to the rate of change of the total mass contained in massive
haloes. 
We obtain this rate from the Press-Schechter formalism for a Gaussian
PDF as
\begin{equation}
\dot F(m,
z)=-\sqrt{\frac{2}{\pi}}\,\frac{\dot{z}\,\delta_\mathrm{c}\,
  (1+z)^3}{\sigma(m)d_{+}(z)}\exp\left(
\frac{-\delta^{2}_{\mathrm{c}}}{2\sigma^2(m)}\right) \frac{\partial\log
  d_+}{\partial\log a},   
\end{equation} 
where an overdot denotes derivative with respect to the cosmic time, a
prime denotes derivative with respect to the redshift, and $F$ is the
fraction of haloes with mass greater than $m$ \citep{1974ApJ...187..425P}.
The critical density for spherical collapse is symbolized by
$\delta_\mathrm{c}$, and $\sigma^2(m)$ is the variance in the initial
density fluctuation field when smoothed with a top-hat filter of a
scale corresponding to mass $m$.  
The rate of growth for perturbations in the linear theory
is denoted by $d_+(z)$. 

The total amount of baryons added to haloes of mass greater than
$10^8$ M$_\odot$ is taken to be the amount of gas available for star
formation.
Gas already present in haloes is not considered for star formation.
We do not consider the contribution of minihaloes as these do not contribute
significantly to the total star formation due to radiative feedback
\citep{2009arXiv0901.0711T}.  
We define the efficiency of star formation, $f_*$ as the fraction of
this gas that is converted into stars. 
The rate of change of stellar mass can now be written as
\begin{equation}
\dot\rho_*(z)=f_*\Omega_\mathrm{b}\rho_\mathrm{c}\dot
F(10^8\,\mathrm{M}_\odot, z),
\label{rhodot}
\end{equation}
where $\rho_\mathrm{c}$ is the critical density and
$\Omega_\mathrm{b}$ is the density parameter for baryons.
Here we have ignored mass lost by stars through winds, outflows, and
supernovae.
This can be taken into account using, for example, population
synthesis models\footnote{A starburst with a Kroupa IMF and an initial
metallicity of 0.02 loses about 10\% of its mass to the ISM through these
effects under normal assumptions.}.
Equation (\ref{rhodot}) illustrates the small number of parameters and
approximations that go into estimating $\dot\rho_*$ in our
formulation. 

Observations of absorption systems in quasar spectra have been used to 
put constraints on the average metallicity of the IGM
\citep{1995AJ....109.1522C, 1997ApJ...490L...1S, 2000AJ....120.1175E,
  2004ApJ...606...92S,2008arXiv0812.2856B,2009arXiv0902.1991R}.
These observations indicate that the amount of C~IV in the IGM does not evolve
significantly between $2 \leq z \leq 5.5$ 
\citep{2001ApJ...561L.153S,2008arXiv0812.2856B,2009arXiv0902.1991R}.
It is not yet clear whether the IGM has been contaminated by metals
throughout, or if the enriched regions of the IGM are restricted to
the neighborhood of galaxies and filaments. 
There are also issues related with understanding the ionization state to map
from absorption by a particular species to metallicity
\citep{2003ApJ...596..768S}. 
Observations also support a correlation between density and metallicity of
the IGM, indicating that regions in proximity of galaxies are enriched
to higher level than regions of IGM far away from galaxies
\citep{2003ApJ...596..768S, 2006ApJ...638...45P, 2006MNRAS.365..615S}.
We assume that the IGM is uniformly enriched at the level indicated by
\citet{2001ApJ...561L.153S} at $z \sim 5.5$.

Winds and outflows are expected to be the dominant processes that lead
to ejection of metals from the inter-stellar medium (ISM) of galaxies.
There is considerable evidence in favor of this mechanism as
observations have detected outflows around almost all galaxies at high
redshifts \citep{2001ApJ...554..981P,2002ApJ...568..558F}.  
The fraction of processed metals that can be deposited from the ISM to
the IGM without disturbing the IGM in an observable manner is not
known.  
Several authors often assume that around $1\%$ of metals produced in
galaxies can be ejected and deposited in the IGM. 
This may also be computed from first principles in detailed
models
\citep{2004ApJ...617..693D,2006ApJ...647..773D,2008MNRAS.385..783S}.  
In these models supernova-driven outflows are responsible for the IGM
enrichment.  
The efficiency of these outflows depends on the star formation
efficiency, the IMF and the efficiency of winds
\citep{2001ApJ...555...92M, 2002ApJ...574..590S, 2005ApJ...624L...1S,
  2003ApJ...588...18F}.  
For a star formation efficiency of 10\% the volume filling factor of
the ejecta can be 20--30\% and at $z\sim 3$ the IGM metallicity could
be around [-3] as detected by \citet{2001ApJ...561L.153S}.

The amount of metals produced per baryon in stars can be computed once
we fix the initial mass function (IMF) of stars, and metallicity of
the star-forming gas \citep{1999ApJS..123....3L,2005ApJ...621..695V}. 
We can write
\begin{equation}
n_Z=f_*f_\mathrm{esc, Z}\,\Omega_\mathrm{b}\frac{\rho_\mathrm{c}}{m_p}\dot
F(10^8\,\mathrm{M}_\odot, z)\,p
\end{equation}
for the number density of metals that reaches the IGM. 
Here $f_\mathrm{esc, Z}$ is the fraction of total metals produced that is
deposited in the IGM and $p$ is the metal yield of the stars per baryon.
Note that we assume the same escape fraction for metals from galaxies of
different masses, whereas it is far more likely that low mass galaxies lose
nearly all the metals from the ISM and more massive galaxies lose
very little \citep{1986ApJ...303...39D}.

Finally, observations of the temperature and polarization anisotropies in
the CMB provide a constraint on the EoR. 
Free electrons produced during reionization scatter the CMB photons
and suppress temperature anisotropies on scales smaller than the
Hubble radius at that time. 
The suppression is of the form
$\mathcal{C}_l^{T^\prime}\!=e^{-2\tau}\mathcal{C}_l^T$ where 
\begin{equation}
\tau =\int n_e(t)\,\sigma_T\,c\,dt
\end{equation}
is the Thomson scattering optical depth.
Here $n_e(t)$ is the number density of free electrons and $\sigma_T$
is the Thomson scattering cross-section.
This damping is degenerate with the amplitude of the primordial power
spectrum. 
The degeneracy is broken by detection of a polarization anisotropy for
scales greater than the Hubble radius at the reionization redshift, an effect
that dominates at scale of the Hubble radius at EoR and has an amplitude
proportional to $\tau$. 
Any model of reionization must reproduce the observed value of optical depth.
For a forecast on anticipated improvements of observational estimation of
$\tau$, please see \citep{2009NewA...14..269C}.

\section{Reionization Model}
\label{reionization}

The reionization history depends on the star formation history of the
universe, which in the simplified models is closely related to the halo
formation history.  
The IMF of stars and the escape fraction for ionizing photons then
give us the number of ionizing photons that are available as a
function of time.  
These can then be used to compute the evolution of the neutral or
ionized fraction of gas in the universe.  
As mentioned above, we assume that star formation is triggered during
formation of haloes.  
As most time scales of interest are longer than the dynamical time scale
over which the bulk of star formation takes place, we assume star
formation to be instantaneous in our model.

\begin{figure*}
\begin{center}
\begin{tabular}{ccc}
\includegraphics[width=2.1truein]{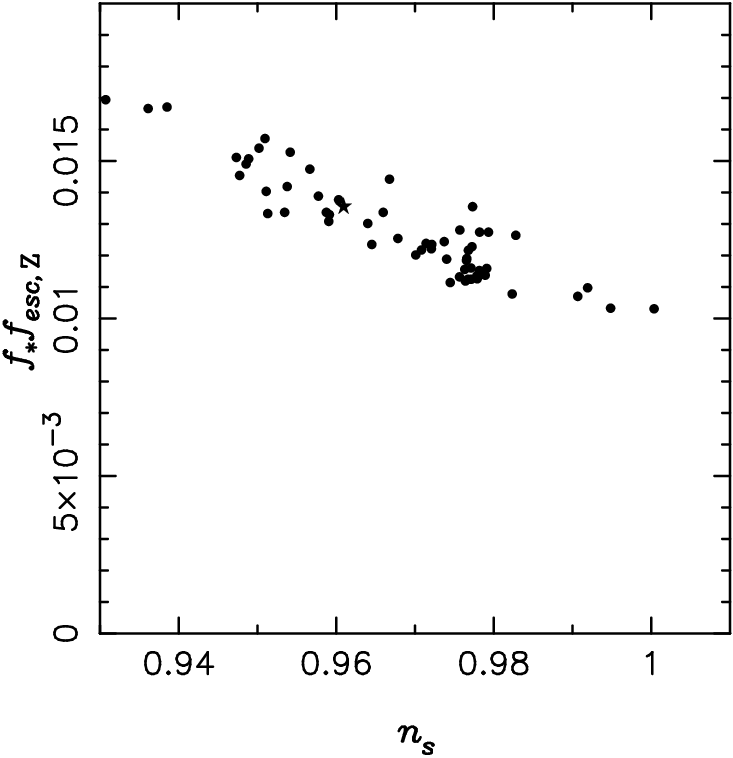} &
\includegraphics[width=2.1truein]{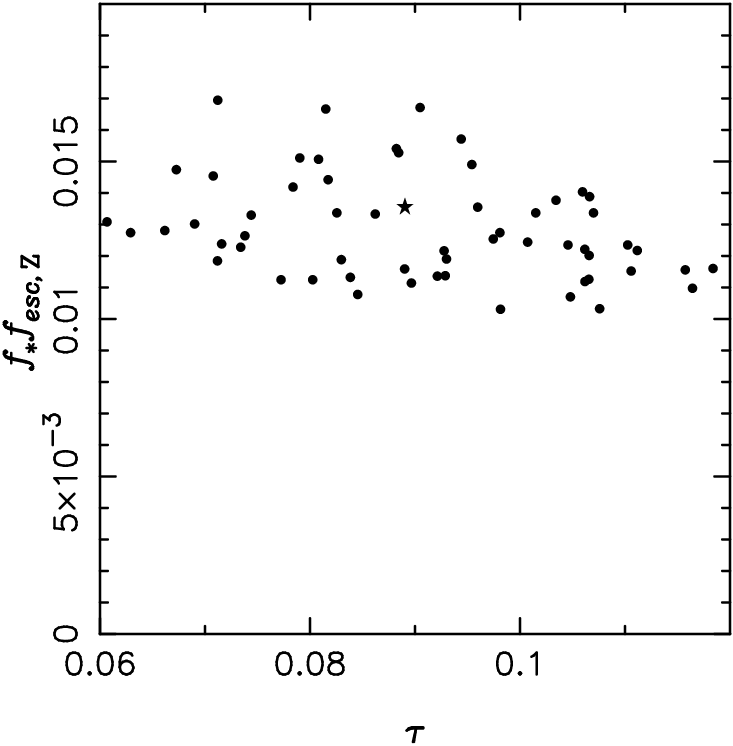} &
\includegraphics[width=2.1truein]{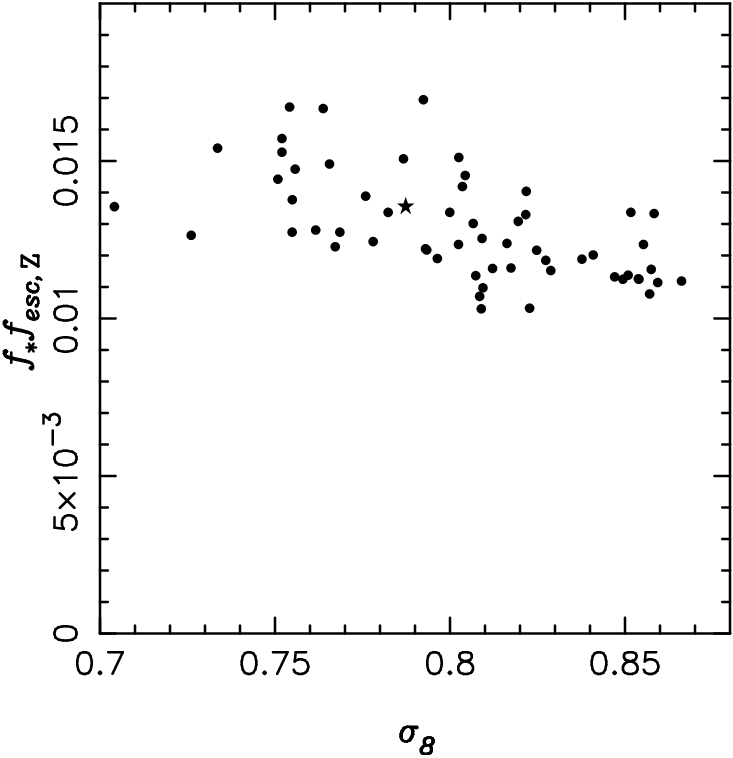} \\
\includegraphics[width=2.1truein]{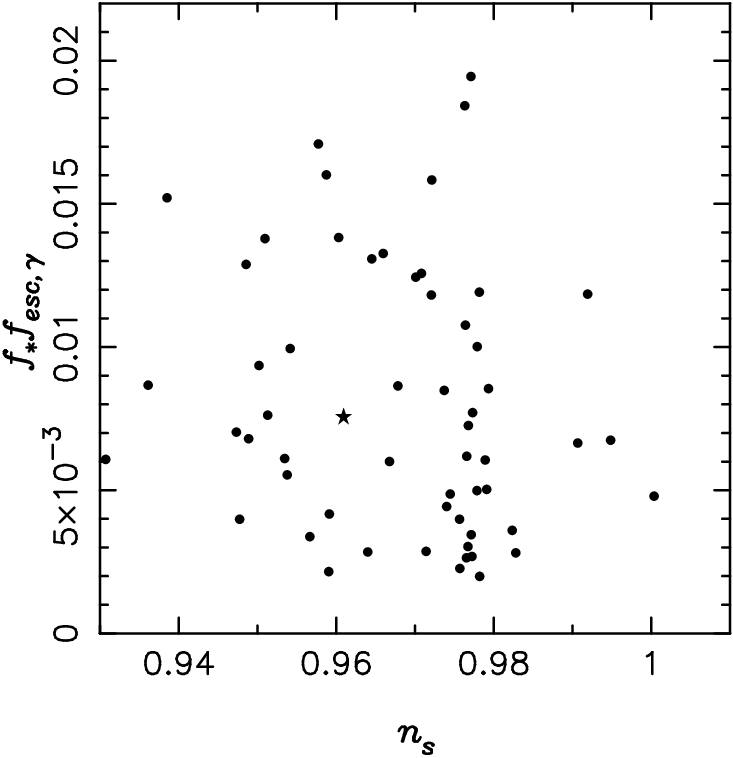} &
\includegraphics[width=2.1truein]{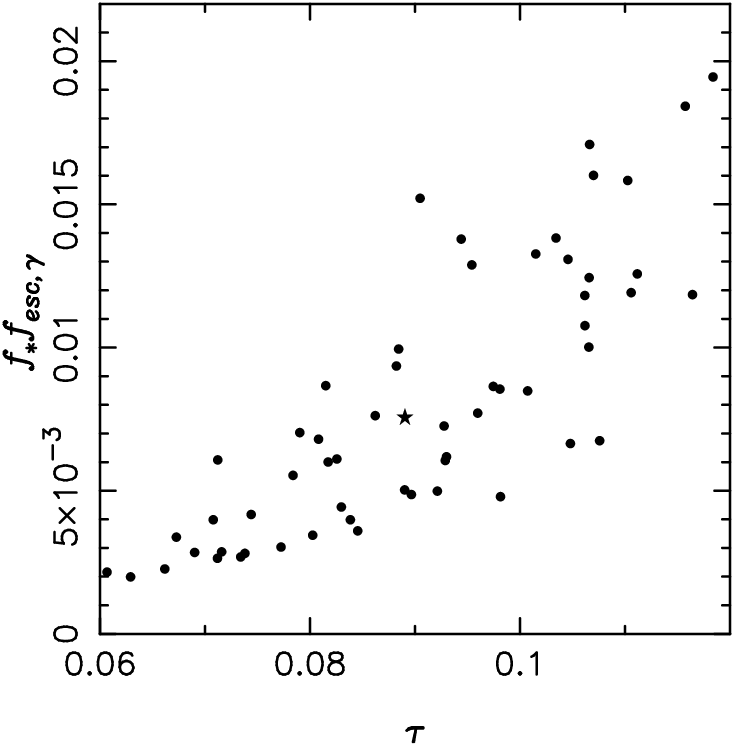} &
\includegraphics[width=2.1truein]{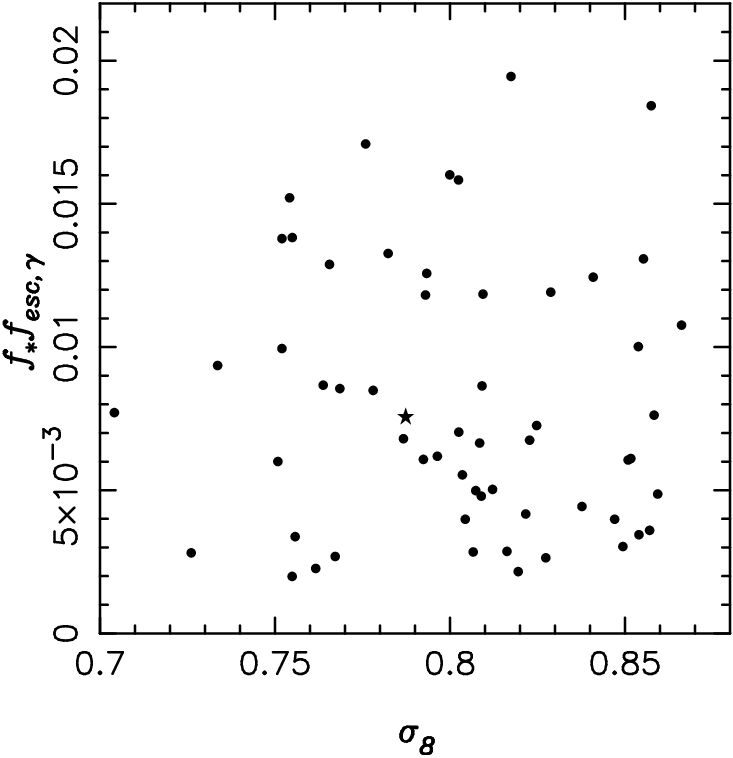} \\
\end{tabular}
\end{center}
\caption{The products $f_{\ast}f_\mathrm{esc, Z}$ (top row) and
  $f_{\ast}f_{\mathrm{esc}, \gamma}$ (bottom row) against cosmological
  parameters for model 1 of Table (\ref{sfmodels}).  The star symbol denotes
  the WMAP5 best fit model.  These points also represent lower bounds
  on $f_\mathrm{esc, Z}$ and $f_{\mathrm{esc}, \gamma}$.  See text for
  details.}
\label{row8fig}
\end{figure*}

We consider a global averaged evolution of ionized fraction instead of
following evolution of HII regions around haloes, the approach used in
most studies \citep{2000ApJ...534..507C,
2005MNRAS.363..818S}. 
Further, we assume that during reionization, a region is either
neutral or completely ionized.  
With these assumptions, the evolution of the ionized fraction evolves
as:
\begin{eqnarray}
\dot{x} &=& -\alpha_{\mathrm{B}}{\mathcal C}^2n_{\mathrm{H}}x +
\sigma_{p}yn_{\mathrm{H}}c(1-x)\label{xtoy} \\
\dot{y} &=& -\sigma_{p}y\,n_{\mathrm{H}}\,c(1-x) +
m_{p}f_{\ast}f_{\mathrm{esc},\gamma}\dot{F}N_{\gamma}.\label{ytoy} 
\end{eqnarray}

Here $x$ is the fractional volume that is ionized, and $y$ is the number of
ionizing photons per baryon.  
$\sigma_{p}$ denotes the effective cross-section of photoionization,
$\alpha_{B}$ is the recombination coefficient for all levels except the ground
state of neutral hydrogen, and $m_p$ denotes the mass of a proton.

The first term on the right hand side of equation (\ref{xtoy})
describes recombination. 
${\mathcal C}$ is the clumping factor defined as ${\mathcal C}^2 =
{\langle}n_{H}^2{\rangle}/{\langle}n_H{\rangle}^2$. 
This term usually involves square of the ionized fraction but in our model we
assume that the ionized fraction is either unity or zero. 
This, when used in volume averaging over the universe with an additional
assumption that the clumping is the same in ionized and neutral regions, leads
to a linear dependence.  
In the process of averaging, the meaning of $x$ changes from the ionized
fraction to the volume filling fraction of the ionized regions.
We can express this in terms of equations:
\begin{equation}
\label{eqls}
\frac{1}{V}{\int}n_{H}^2 x^2 dV = \frac{{\langle}n_{H}^2{\rangle}}{V}{\int}
x^2 dV = \frac{{\langle}n_{H}^2{\rangle}}{V}{\int} x dV =
{\langle}n_{H}^2{\rangle}x.
\end{equation}
We have assumed that the clumping factor is the same in all parts of
the universe, this allows us to take $\langle n_{H}^2\rangle$ outside the
integral.  
The third equality in equation (\ref{eqls}) follows from the
definition of $x$ as a filling fraction. 
We also expect ${\mathcal C}$ to change with the evolution of clustering.  
We take this dependence to be of the form \citep{2007MNRAS.376..534I}
\begin{equation}
{\mathcal C}^2 \simeq 26.2917\exp\left[-0.1822z + 0.003505z^2\right].
\end{equation}

Sources of ionizing radiation are represented in the last term of
equation~(\ref{ytoy}), $\dot{F}$ being related to the formation rate
of collapsed haloes. 
This is obtained from the Press-Schechter formalism as described
above.
$N_{\gamma}$ denotes the number of photons produced per unit mass of
star formation.
Ionization of neutral hydrogen is described in the last term on the
right hand side of equation~(\ref{xtoy}).
This term occurs in both equations.  
We neglect the contribution of collisional ionization.

\begin{figure*}
\begin{center}
\begin{tabular}{ccc}
\includegraphics[width=2.1truein]{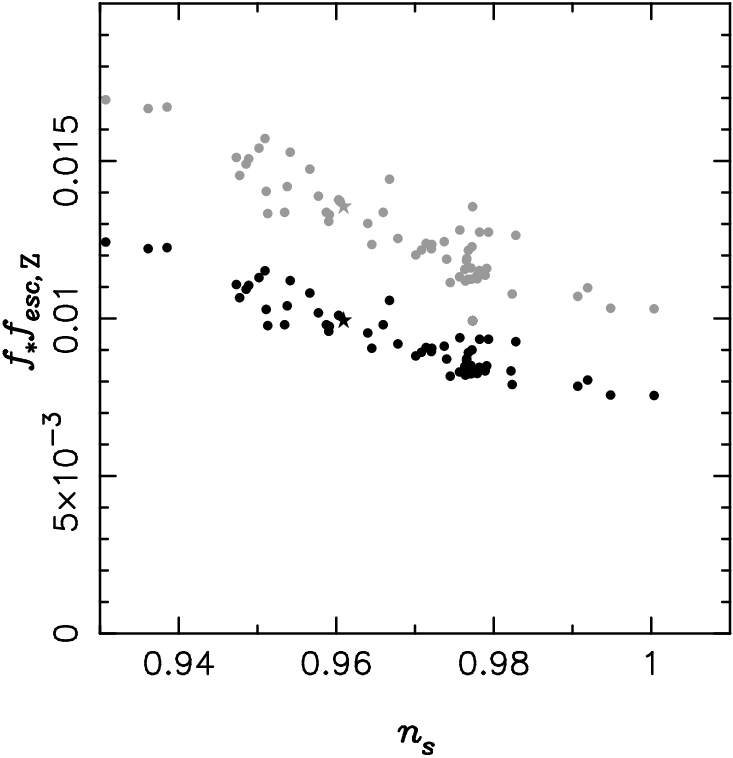} &
\includegraphics[width=2.1truein]{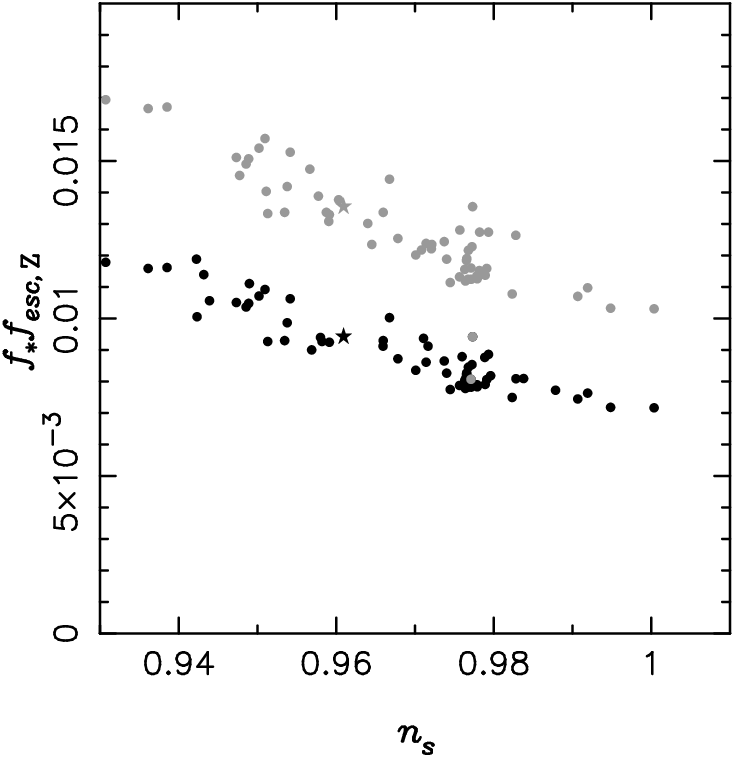} &
\includegraphics[width=2.1truein]{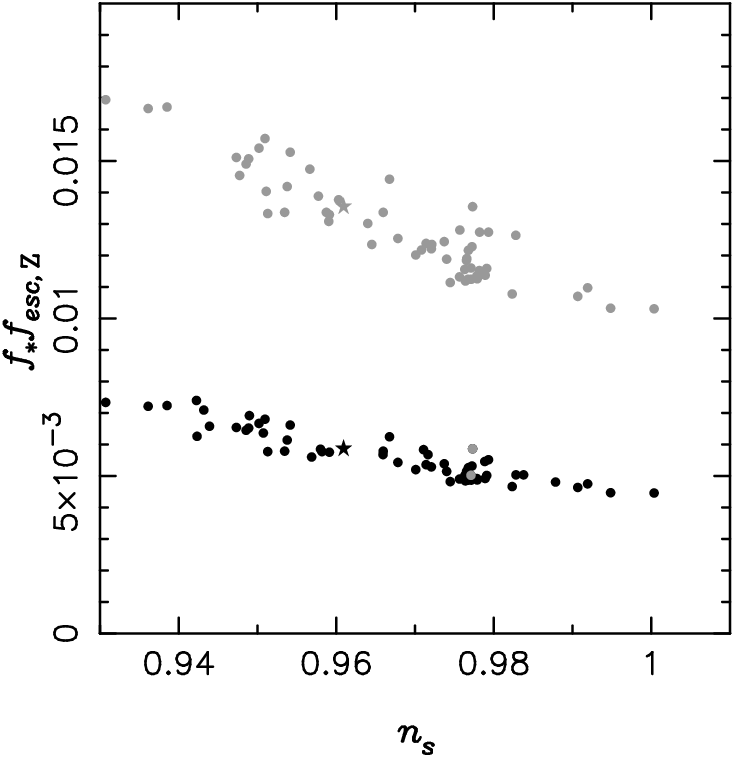} \\
\includegraphics[width=2.1truein]{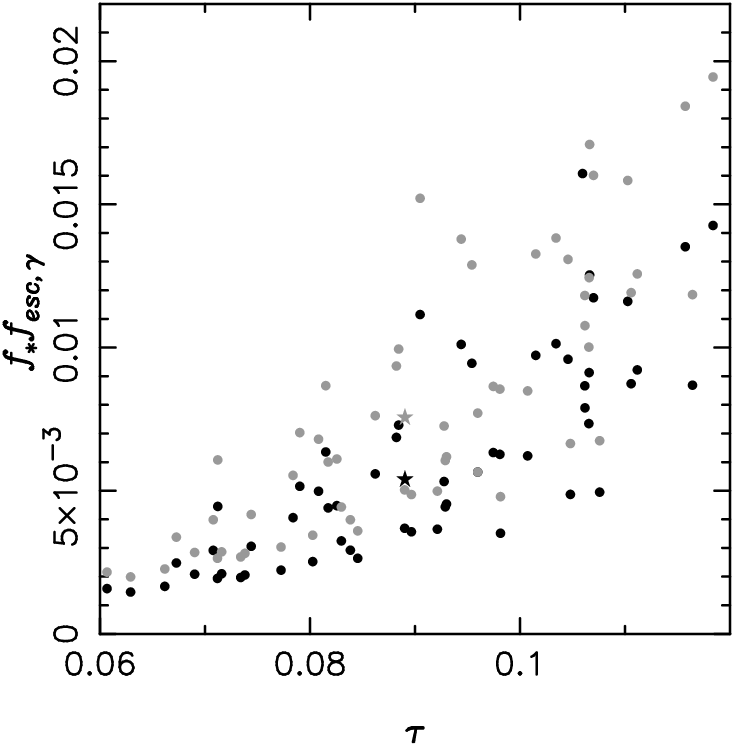} &
\includegraphics[width=2.1truein]{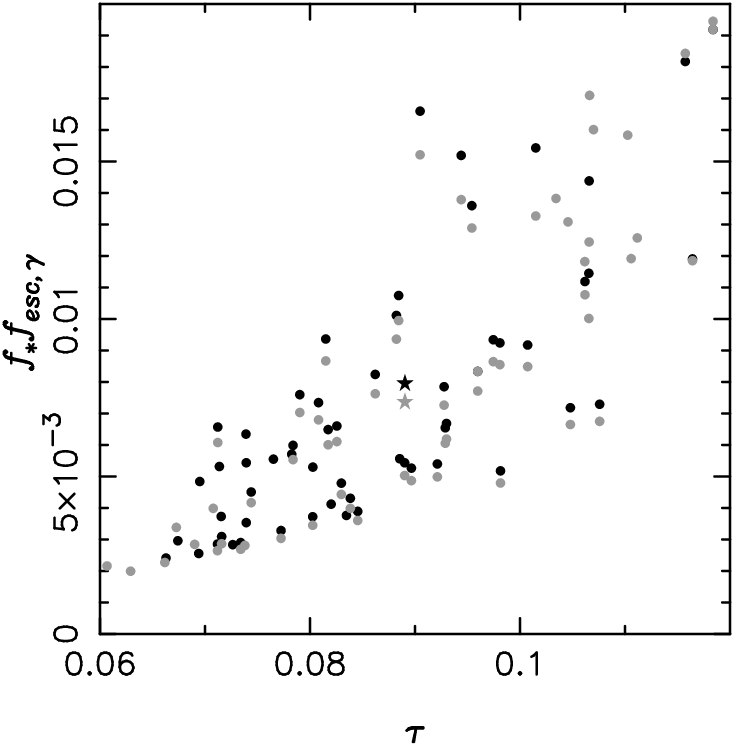} &
\includegraphics[width=2.1truein]{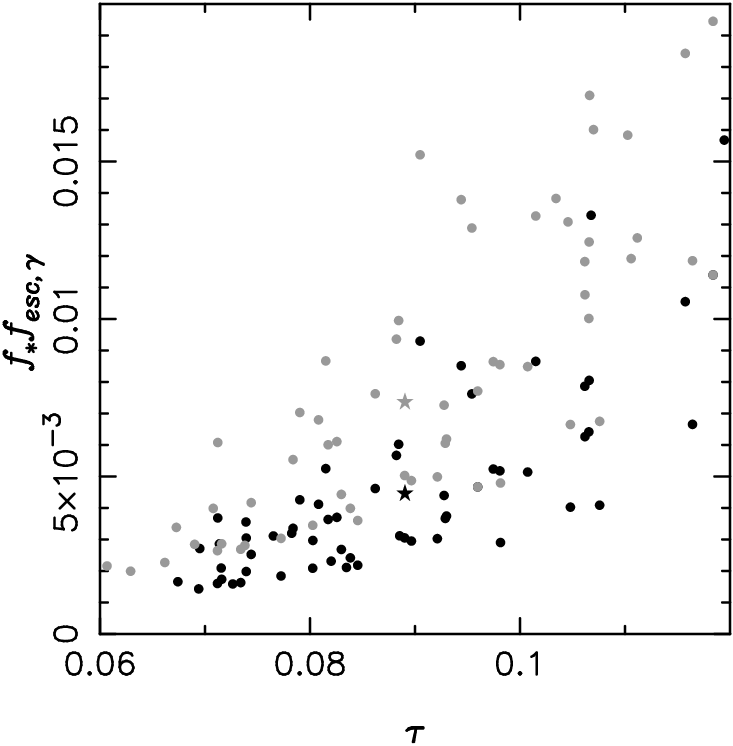} \\
\end{tabular}
\end{center}
\caption{The products $f_{\ast}f_\mathrm{esc, Z}$ and
 $f_{\ast}f_{\mathrm{esc}, \gamma}$ for model 2 (left column), model 3
 (center column) and model 4 (right column) of Table (\ref{sfmodels}).
 Corresponding plots from Figure (\ref{row8fig}) are superimposed in
 grey.  Filled stars denote values for the WMAP5 best-fit models.
 These points are represent lower bounds on $f_\mathrm{esc, Z}$ and
 $f_{\mathrm{esc}, \gamma}$.  See text for details.}
\label{otherfigs}
\end{figure*}

We solve these equations numerically for different cosmological models. 
The system of equations (\ref{xtoy}) and (\ref{ytoy}) is ``stiff,'' since
$(x=0,y=0)$ is a stable point and time scales for evolution of $x$ and $y$ are
very different.  
Further, $x$ is bounded from above (by unity) while $y$ is not. 
Thus the usual forward differencing methods do not give accurate solutions
easily. 
We bypass this problem by noting that during the process of reionization
almost every ionizing photon will be immediately absorbed by the
medium\footnote{There are two approximations being discussed here: one
  approximation is that all UV photons are available for ionizing atoms.  This
  is not really true as ionized regions can be expected to host an ionizing
  background.  The other approximation is in solving the equations that we
  arrive at with the first approximation.}. 
This means that the two terms in the right hand side of the second equation
are of the same order till $x$ becomes nearly equal to $1$, whereas the left
hand side is much smaller and may be assumed to be zero.  
This reduces the system of equations to a single equation, which can now be
solved using forward differencing methods. 
Note that this approximation is not valid when $x$ approaches 1, although in
practice the approximate solution is fairly accurate up to $x \sim 0.9$. 
Indeed, if we use the approximation up to $x=1.0$ then we make an error in
estimation of $\tau$ of less than $5\%$.

We take $\alpha_{\mathrm B} = 1.0 \times 10^{-13} \; \mathrm{cm^3} \,
\mathrm{sec^{-1}}$, ignoring its dependence on temperature.
This dependence is fairly weak at temperatures of interest. 
We use $\sigma_p = 6.30 \times 10^{-18}\;\mathrm{cm^2}$. We thus
assume that most of the ionizing radiation is around the Lyman limit. 
The number of ionizing photons released per baryon of stars formed,
denoted by $N_{\gamma}$, depends on the initial mass function (IMF) of
the stars.
We obtain this number from the {\scshape starburst99} stellar
population synthesis code \footnote{We consider instantaneous
starbursts with a fixed total stelar mass of $10^6$ M$_\odot$.  We use
the Geneva evolutionary tracks for models with metallicity $0.001$ and
the Padova tracks with AGB stars for models with metallicity
$0.0004$.} \citep{1999ApJS..123....3L, 2005ApJ...621..695V}.

\section{Discussion} 
\label{results}

Our aim here is to study a variety of models with varying cosmological
parameters as well as parameters related to star formation and
enrichment.
We consider a random subset of flat $\Lambda$CDM models allowed by
WMAP5 \citep{2008arXiv0803.0547K, 2008arXiv0803.0586D}.
We do not consider models with massive neutrinos or a non-vanishing tensor
component, or models where the primordial power spectrum deviates from a pure
power law.  
We use only WMAP constraints for limiting cosmological parameters.
We used the MCMC chains made available by the WMAP team
(\url{http://lambda.gsfc.nasa.gov/}). 
We considered a random subset of all models allowed with a confidence level of
$68\%$ from the MCMC chains.
We studied a handful of models for parameters related to star
formation; details of these are given in Table (\ref{sfmodels}).
The table lists the IMFs used in our study\footnote{We should note that there
  is considerable uncertainty in the shape of the IMF in the local
  neighborhood \citep{2002ASPC..285...86K,2008arXiv0809.4261C}.  This can
  easily have a significant impact on our conclusions.  The uncertainties
  introduced by other assumptions and approximations should be seen with the
  uncertainty in the IMF as the reference.}. 
We have also listed the amount of ionizing photons produced per baryon
in stars, and the total metal yield per baryon for these IMFs.
These numbers also depend on the metallicity of gas from which stars
form and this is listed in the table as well.
It is interesting to note that for a given IMF, as the metallicity
increases, the production of ionizing photons per baryon comes down
but the amount of enriched material returned to the ISM increases.
Gas that forms the first stars is likely to have primordial abundance
\citep{2000PhR...333..389O}.
In our analysis of all the models, we keep metallicity fixed and hence it is
appropriate to use low values of input metallicity. 

Metallicity of the IGM constrains the product $f_\ast f_\mathrm{esc,Z}$ for a
given model. 
Similarly we constrain the product $f_\ast f_{\mathrm{esc},\gamma}$
with the optical depth due to reionization for the CMB. 
Figure (\ref{row8fig}) shows these products for all cosmological
models studies here, when the star formation parameters for model~1 in
Table~(\ref{sfmodels}) are used.
Given that the efficiency of star formation can at best be $100\%$,
i.e., $f_\ast \leq 1$, the points in Figure (\ref{row8fig}) also
represent lower bounds on $f_\mathrm{esc,Z}$ and
$f_{\mathrm{esc},\gamma}$.
These are shown as a function of the slope of the primordial power
spectrum ($n_s$), optical depth due to reionization
($\tau_{\scriptscriptstyle\mathrm{CMB}}$),
and, amplitude of clustering at the scale of $8$ h$^{-1}$Mpc
($\sigma_8$). 
The best fit WMAP5 model is marked in each panel as a star.  
We find that there is some correlation between the lower bound on
$f_{\ast}f_{\mathrm{esc},\gamma}$ and $\tau_{\scriptscriptstyle\mathrm{CMB}}$,
and also between $f_{\ast}f_\mathrm{esc,Z}$ and $n_s$.
There are weak correlations with other cosmological parameters but nothing
as remarkable as the two mentioned above.

The efficiency of star formation is likely to be much less than unity in any
realistic scenario.
Indeed, if we try to keep the different efficiencies and escape fractions at
the same order then we require these to be around $0.1-0.15$, or $10-15\%$.  
While the escape fraction for ionizing photons and star formation efficiency
we have obtained are comparable to those found in other studies, these are
higher than the values seen in local galaxies.
In particular, it is not clear if it is possible to expel $10\%$ of
the metals from the ISM to distant parts of the IGM using known physical
mechanisms. 
Before commenting on the numbers, let us consider the sensitivity of the
result to our assumptions. 
\begin{itemize}
\item
If the IMF has a low mass cutoff that is higher than the $0.1$~M$_\odot$ used
for model~1 from Table~(\ref{sfmodels}), then a larger fraction of mass goes
into high mass stars that produce the ionizing photon flux and the enriched
material.   
This can lower the required $f_\mathrm{esc,Z}$ by a significant amount. 
The product $f_{\ast}f_\mathrm{esc,Z}$ for model~2 from Table~(\ref{sfmodels})
is shown in Figure~(\ref{otherfigs}).
There is some evidence that there are more intermediate mass stars in the
population of metal poor stars in the halo of the Galaxy as compared to metal
rich stars, if we normalize the two distributions at low stellar
masses \citep{2007ApJ...665.1361T,2007ApJ...658..367K}. 
Thus a higher cutoff for $M_{low}$ may be required for explaining other
observations.  
\item
Our analysis assumes that the metallicity of gas that forms stars is fixed.
If we do a self-consistent analysis where this is allowed to evolve,
the gas metallicity gradually increases. 
It is then clear from Table~(\ref{sfmodels}) that later generations of stars
will enrich the ISM faster.
As an illustration, we can see the results of analysis with higher fixed input
metallicity for models~2--4 in Figure~(\ref{otherfigs}). 
Our estimates show that $f_\mathrm{esc,Z}$ required to satisfy observations
can come down by a few tens of percents due to this.
There is a corresponding increase in $f_\mathrm{esc,\gamma}$ due to stars with
higher metallicity producing fewer ionizing photons.
\item
We have assumed that haloes with mass above $10^8$ M$_\odot$ can form stars.
The standard approach is to assume that radiative feedback during the EoR
increase this by about an order of magnitude in regions that have been
photo-ionized \citep{1992MNRAS.256P..43E}.
We do not take this into account as it has been pointed out that the actual
effect may only be to reduce the efficiency of star formation in lower mass
haloes \citep{2008MNRAS.390.1071M}. 
If we do consider the effect of radiative feedback as disabling star formation
then it leads to a reduction in total gas available for star formation, and
hence requires slightly higher efficiencies and escape fractions.  
We find that this effect requires an increase in the two products by about
$20\%$. 
\item
We have assumed that the universe is enriched uniformly.  
It may very well happen that the enrichment process is effective only in the
vicinity of galaxies. 
In such a case only overdense regions are enriched. 
The escape fraction of metals required can be lowered by as much as a factor
of two if this is the case. 
\end{itemize}
Thus we may require only around $5\%$ of the ISM to be ejected to the IGM on
an average in models with $f_\ast \simeq 0.2$
This is comparable with semi-analytic \textit{ab initio} models of
early star formation, outflows and IGM enrichment that have been
studied in the literature.  

We have assumed the same loss fraction for ISM for galaxies over the entire
range of masses. 
This, of course, is not true.
We expect that the low mass galaxies can potentially disperse a large
fraction of the ISM in supernova explosions but
larger galaxies can retain most of their ISM \citep{1974MNRAS.169..229L,
  1986ApJ...303...39D}. 
If most of the IGM enrichment is done by metals that form in dwarf
galaxies then the constraint is not very stringent. 
In most models, the fraction of mass in galaxies with a halo mass of less than
$10^{10}$~$M_\odot$ is larger than $10\%$ even at $z \simeq 6$.
If these galaxies lose a significant fraction of the ISM on an average and
heavier galaxies lose very little mass then we can comfortably satisfy the
constraints from enrichment of the IGM. 

\begin{figure}
\includegraphics[width=2.8truein]{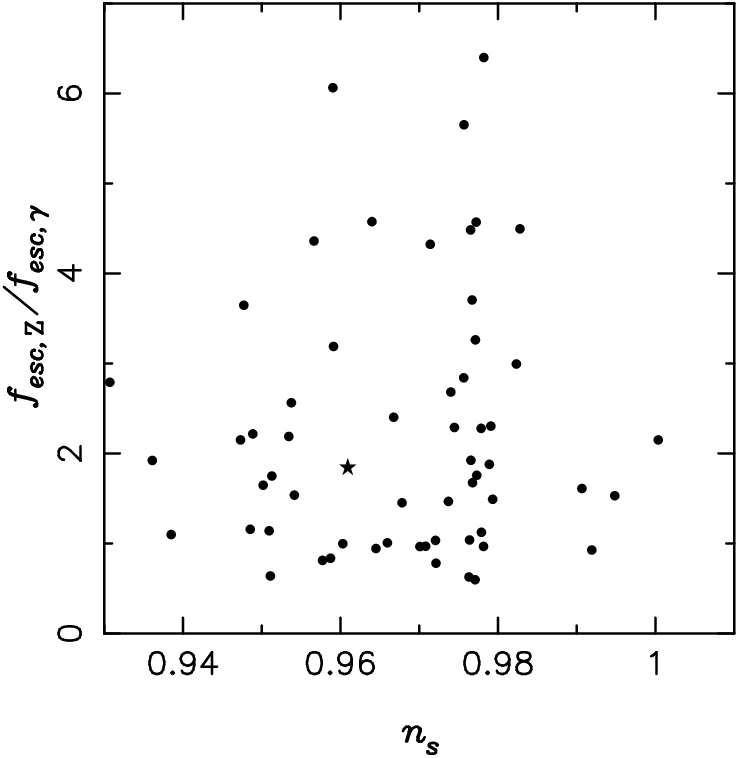}
\caption{Ratio $f_\mathrm{esc,Z}/f_{\mathrm{esc},\gamma}$ for the first IMF
listed in Table~(\ref{sfmodels}). See text for details.}
\label{ratiofigs}
\end{figure}

We now turn our attention to the ratio of escape fractions for photons and
metals. 
Figure~(\ref{ratiofigs}) shows the ratio
$f_\mathrm{esc,Z}/f_{\mathrm{esc},\gamma}$ for the first IMF listed in
Table~(\ref{sfmodels}). 
It is interesting to note that the ratio
$f_\mathrm{esc,Z}/f_{\mathrm{esc},\gamma}$ is of order unity, differing from
unity by at most a factor of a few.  
Thus the fraction of ionizing photons that escape galaxies is broadly of the
same order as the fraction of metals that must leak into the IGM in order to
explain the observed enrichment of the IGM.
We have not plotted this ratio for other IMFs in Table~(\ref{sfmodels}) as the
expected change can be seen from Figure~2. 
Indeed, the change in the ratio is less than a factor two as we consider IMFs
listed in rows $2-4$ of the Table~(\ref{sfmodels}).
The last row in the Table~(\ref{sfmodels}) is more appropriate for late time
star formation and we need not discuss that here. 

Most studies of reionization have tended to focus on the escape of ionizing
photons. 
In order to satisfy observations of $\tau$ or the luminosity function of high 
redshift galaxies, these often invoke a top heavy IMF for the first generation
of stars \citep{2003ApJ...591L...5C, 2003ApJ...595....1H,
  2003ApJ...588L..69W, 2004PASP..116..103B}. 
Other sources of ionizing radiation like AGNs, magnetic fields and decaying
dark matter particles have also been studied
\citep{2004PhRvL..92c1301P,2008PhRvD..78h3005S}.  
It is interesting to note that all such modifications lead to an enhanced
production of ionizing photons without affecting the production of metals in a
significant manner. 
This is because the very massive stars are expected to implode and do not
enrich the ISM with the products of nuclear fusion that takes place in the
core. 
The ratio 
$f_\mathrm{esc,Z}/f_{\mathrm{esc},\gamma}$ changes on addition of extra
sources of ionizing radiation. 
In view of the arguments presented above, all modifications that have been
discussed so far lead to a smaller $f_{\mathrm{esc},\gamma}$. 
In other words, the ratio plotted in Figure~(3) should be thought of as a lower
bound.  

An important implication of this is that the constraints from enrichment of
the IGM require certain amount of star formation, and this requirement needs
to be satisfied even when we invoke other sources of ionizing radiation during
the epoch of reionization.  
That is, adding new potential sources of ionizing radiation can be helpful
only in lowering the escape fraction of ionizing photons and not in lowering
the amount of star formation\footnote{An exception is the scenario where the
  universe is not enriched throughout.  In such a case even $f_\mathrm{esc,Z}$
  can be reduced by a significant amount.}.
We may even end up with a scenario where a much larger fraction of processed
elements need to be transferred from the ISM to the IGM as compared to the
fraction of UV photons escaping from galaxies.

\section{Conclusions}

We have compared simple models for star formation in the early universe with
two observational constraints. 
The simplicity of the model allows us to consider variation in cosmological
parameters as well.
We present a summary of our results here:
\begin{itemize}
\item
The product of star formation efficiency and escape fraction of ionizing
photons, $f_{\ast}f_{\mathrm{esc},\gamma}$, is correlated with the optical
depth due to reionization.
\item
The product of star formation efficiency and escape fraction of ISM,
$f_{\ast}f_{\mathrm{esc},Z}$, is anti-correlated with the index of the
primordial power spectrum.
\item
These are weak correlations, in the sense that the values for the two products
do not change strongly for small changes in the cosmological parameter in
question. 
\item
We do not find any other correlation amongst parameters of star formation and
cosmological parameters.
\item
We are able to satisfy observational constraints with the standard initial
mass function for stars observed in the local universe
\citep{2002ASPC..285...86K}, and with reasonable values for star formation
efficiency and escape fractions for photons and ISM.  
Given that the local IMF itself is somewhat ill constrained, this implies that
we do not require a significant evolution of the IMF in order to explain
observations considered here.
\item
Small variations in the IMF, indicated by observations of metal poor stars in
the Galaxy, reduce the efficiency of star formation and the escape fractions
required for the standard IMF.
\item
Approximations used by us in the model do not change the overall numbers by
more than $10-20\%$.  
Indeed, different approximations change numbers in different directions so we
can consider the overall results to be fairly robust.
\item
Our model allows us to estimate the ratio of the two escape fractions. 
We find that the two escape fractions are of the same order. 
\item
If we consider other potential sources of ionizing photons then the required
escape fraction for photons can come down, however the escape fraction for
processed elements does not change.  
Indirectly, the required amount of star formation is required to remain the
same unless there is some very efficient mechanism for transporting processed
elements into the IGM while keeping the escape fraction of ionizing photons
low. 
\end{itemize}

The most important conclusion of this work is that star formation without a
significant evolution of the IMF is sufficient for satisfying the two
constraints considered here. 
The escape fractions, and/or the star formation efficiency is required to be
higher than we see in local galaxies.
One can consider other sources of ionizing radiation, indeed at least some of
these must be present.
But as we have pointed out, these help in reducing only the escape fraction
for ionizing radiation as none of the other potential sources help in
transporting enriched material from the inter-stellar medium to the
inter-galactic medium.
This highlights the significance of the constraint arising from enrichment of
the IGM for epoch of reionization studies.

\section*{Acknowledgments}

The authors would like to thank Shiv Sethi, K. Subramanian, Saumyadip
Samui, R. Srianand and T. Sivarani for useful discussions and
comments.  
GK acknowledges helpful discussions with H.~K.~Jassal and T.~Roy
Choudhury.
Computational work for this study was carried out at the cluster
computing facility in the Harish-Chandra Research Institute
(\url{http://cluster.hri.res.in}).  
This research has made use of NASA's Astrophysics Data System. 
We acknowledge the use of the Legacy Archive for Microwave Background
Data Analysis (LAMBDA). 
Support for LAMBDA is provided by the NASA Office of Space Science.

\label{lastpage}
\end{document}